\documentclass{pasj00}
\draft

\begin{document}
\SetRunningHead{A. Nakajima et al.}{Clustering of low-luminosity star-forming galaxies}
\Received{yyyy/mm/dd}
\Accepted{yyyy/mm/dd}

\title{Clustering Properties of Low-Luminosity Star-Forming galaxies at $z = 0.24$ and 0.40 in the Subaru Deep Field
\footnotemark[*]}

\author{
        Aki \textsc{Nakajima},\altaffilmark{1}
        Yasuhiro \textsc{Shioya},\altaffilmark{2}
        Tohru \textsc{Nagao},\altaffilmark{1,2}
        Tomoki \textsc{Saito},\altaffilmark{2}
        Takashi \textsc{Murayama},\altaffilmark{3}
        Shunji S. \textsc{Sasaki},\altaffilmark{1,3}
        Asuka \textsc{Yokouchi},\altaffilmark{3} and 
        Yoshiaki \textsc{Taniguchi}\altaffilmark{2}
        }

\email{nakajima@cosmos.phys.sci.ehime-u.ac.jp}

\footnotetext[*]{Based on data collected at 
	Subaru Telescope, which is operated by 
	the National Astronomical Observatory of Japan.}

\altaffiltext{1}{Graduate School of Science and Engineering, Ehime University, 
        Bunkyo-cho, Matsuyama 790-8577}
\altaffiltext{2}{Research Center for Space and Cosmic Evolution, Ehime University, 
        Bunkyo-cho, Matsuyama 790-8577}
\altaffiltext{3}{Astronomical Institute, Graduate School of Science,
        Tohoku University, Aramaki, Aoba, Sendai 980-8578}


\maketitle

\begin{abstract}
We present our analysis on the clustering properties of star-forming galaxies selected 
by narrow-band excesses in the Subaru Deep Field. Specifically we focus on H$\alpha$ 
emitting galaxies at $z = 0.24$ and $z = 0.40$ in the same field, 
to investigate possible evolutionary signatures of clustering properties of 
star-forming galaxies. Based on the analysis on 228 H$\alpha$ emitting galaxies with 
$39.8 < \log L({\rm H}\alpha) < 40.8$ at $z = 0.40$, 
we find that their two-point correlation function is estimated as $\xi = (r/1.62^{+0.64}_{-0.50}{\rm Mpc})^{-1.84\pm0.08}$. 
This is similar to that of H$\alpha$ emitting galaxies in the same H$\alpha$ luminosity range 
at $z = 0.24$, $\xi = (r/1.88^{+0.60}_{-0.49}{\rm Mpc})^{-1.89 \pm0.07}$.
These correlation lengths are smaller than those for the brighter galaxy sample studied by Meneux et al. (2006) in the same 
redshift range. The evolution of correlation length between $z = 0.24$ and $z = 0.40$ 
is interpreted by the gravitational growth of the dark matter halos. 
\end{abstract}

\section{INTRODUCTION}

Formation and evolution of galaxies and its dependence on the environment, 
e.g., the formation of large-scale structure, are fundamental problems of 
the modern cosmology. 
Studying the evolution of the clustering of galaxies is useful to understand the 
evolution of galaxies and that of large-scale structures. 
Galaxies are considered to form within dark matter halos. 
Owing to the WMAP observations (Spergel et al. 2007), 
the cosmological parameters are well determined and 
the evolution of the clustering of dark matter 
halos is now well understood both analytically (Mo \& White 1996; Sheth \& Tormen 1999) 
and from N-body simulations (Jenkins et al. 1998; Kauffmann et al. 1999; Springel et al. 2005). 
However, physical relationship between galaxy populations and the dark matter halos is not clear, 
since galaxy formation processes include gas cooling, star formation, and feedback. 

At the present universe, galaxies selected from optical observations are considered 
to trace the distribution of underlying dark matter halos (Verde et al. 2002). 
However, the strength of the clustering depends on properties of galaxy populations: 
e.g., luminosity (Zehavi et al. 2005), color or spectral type (Zehavi et al. 2002), and 
stellar mass (Li et al. 2006). 
The dependence of clustering on the properties of galaxies gives constraint on 
theories of galaxy formation. 
To constrain the model of galaxy formation, it is useful to investigate such dependences in detail. 
In this paper, we concentrate the 
dependence of clustering on the luminosity of galaxies. 
Norberg et al. (2001) fit the relative bias $b/b^*$ as a function of $L/L^*$, 
$b/b^* = 0.85 + 0.15 L/L^*$, 
where $b^*$ is the bias for $L^*$ galaxies and the bias is defined as the square root 
of the ratio of the galaxy and dark matter correlation functions, $b \equiv (\xi_{\rm g}/\xi_{\rm DM})^{1/2}$. 
Using the SDSS data, Tegmark et al. (2004) gave another fitting function, 
$b/b^* = 0.85 + 0.15 L/L^* - 0.04(M-M^*)$. 
The relative bias at the low-luminosity end is smaller than the Tegmark's relation 
in Zehavi et al. (2005). Therefore the relative bias of low-luminosity galaxies 
is still controversial. 

Recently, the dependence of clustering on the luminosity is studied 
up to $z \sim 1.5$ from the VIMOS-VLT Deep Survey (Marinoni et al. 2005; Meneux et al. 2006; Pollo et al. 2006), 
the DEEP2 Galaxy Redshift Survey (Coil et al. 2006), 
and the Canada-France Legacy Survey (CFHTLS, McCracken et al. 2008). 
Brighter galaxies are more strongly clustered than low luminosity galaxies and 
the dependence of relative bias on the luminosity at $z \sim 0.8$ 
is in agreement with that in the local Universe (Marinoni et al. 2005). 
The observed decreasing of the clustering strength is consistent with simple gravitational 
growth picture (Meneux et al. 2006). However, the evolution of the clustering of 
low-luminosity galaxies is not clear till now, since those objects are too faint for 
spectroscopic surveys. 

In this paper, we present the clustering properties of H$\alpha$ emitters 
at $z = 0.24$ and $z = 0.40$, 
which are low-luminosity active star-forming galaxies in the Subaru Deep Field. 
Their mean absolute magnitude is $\langle M_B \rangle \sim -17.0$, which is 1 mag 
fainter than the previous observations (Pollo et al. 2006; Meneux et al. 2006). 
We therefore study the evolution of galaxies with low luminosities. 
Using the narrowband imaging survey, we select star-forming galaxies 
within the restricted redshift to fainter magnitudes.
Throughout this paper, magnitudes are given in the AB system.
We adopt a flat universe with $\Omega_{\rm matter} = 0.3$,
$\Omega_{\Lambda} = 0.7$,
and $H_0 = 70 \; {\rm km \; s^{-1} \; Mpc^{-1}}$.

\section{PHOTOMETRIC CATALOG}

In this paper, we use the official photometric catalogs of 
the Subaru Deep Field (SDF) project that is a very deep optical imaging survey using the Suprime-Cam
(Miyazaki et al. 2002) on the 8.2 m Subaru Telescope (Kaifu et al. 2000;
Iye et al. 2004) at Mauna Kea Observatories. 
The SDF is located near the North Galactic Pole, centered at $\alpha$(J2000) 
= $13^{\rm h} 24^{\rm m} 38^{\rm s}.9$ and $\delta$(J2000) = $+27^\circ
29^\prime 25^{\prime\prime}.9$.
Details of the SDF project are given in Kashikawa et al.(2004).
The SDF official photometric catalogs are 
obtained from the SDF web site (http://step.mtk.nao.ac.jp/sdf/project/).
These official catalogs contain 5 broadband ($B,V,R_{c},i^{\prime}$, and $z^{\prime}$)
and 2 narrowband ($NB816$ and $NB921$) photometric data.
In this work, we use the $i^\prime$-selected catalog with $2^{\prime \prime}$ diameter aperture photometry.
The PSF size in this catalog is 0.98 arcsec (Kashikawa et al. 2004).
Since the Galactic extinction is not corrected in the magnitudes in 
the official catalog, we applied the Galactic extinction correction of 
$E(B-V)=0.017$ (Schlegel et al. 1998). 
A photometric correction for each band is 
$A_B=0.067$, $A_V=0.052$, $A_{R_{\rm C}}=0.043$, $A_{i^\prime}=0.033$, 
$A_{z^\prime}=0.025$, $A_{NB816}=0.030$, and $A_{NB921}=0.024$. 

The narrowband filters used in SDF are $NB816$ [$\lambda_c= 8150$ \AA, 
$\Delta \lambda({\rm FWHM}) = 120$ \AA] and $NB921$ [$\lambda_c= 9196$ \AA, 
$\Delta\lambda$(FWHM) = 132 \AA]. These narrowband filter data are used 
for searching H$\alpha$ emitters at $z = 0.24$ and $z = 0.40$; 
note that the H$\alpha$ luminosity function and the angular correlation function 
of H$\alpha$ emitters at $z = 0.24$ in the SDF has been already studied by 
our group (Morioka et al. 2008). 
The limiting magnitudes for a $3\sigma$ detection on a $2^{\prime \prime}$ 
diameter aperture are $B_{\rm lim,3\sigma}=28.45$, $V_{\rm lim,3\sigma}=27.74$, $R_{\rm c \; lim,3\sigma}=27.80$, $i^\prime_{\rm lim,3\sigma}=27.43$, 
$z^\prime_{\rm lim,3\sigma}=26.62$, $NB816_{\rm lim,3\sigma}=26.63$, and $NB921_{\rm lim,3\sigma}=26.54$ .

\section{RESULTS}

\subsection{Selection of H$\alpha$ emitters}

\begin{figure}
\begin{center}
\FigureFile(150mm,150mm){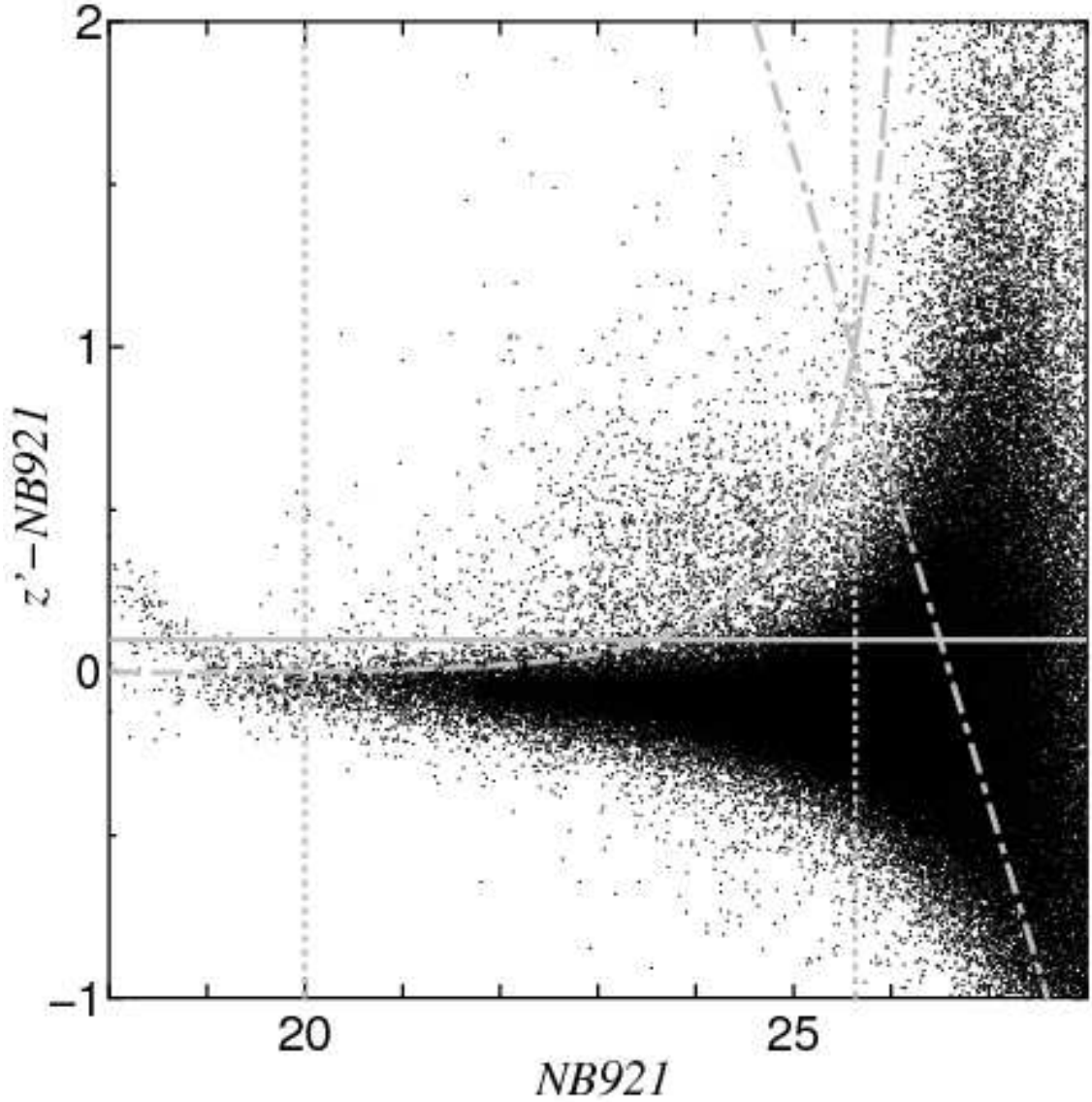}
\end{center}
\caption{Diagram between $z^{\prime}-NB921$ vs. $NB921$. 
The gray horizontal solid line corresponds to $z^{\prime}-NB921=0.1$. 
The gray vertical dotted lines correspond to $NB921 = 20$ and 25.63 (see text).
The gray dashed line shows the distribution of $3\sigma$ error. 
The gray dot-dashed line shows the $3 \sigma$ limiting magnitude of $z^{\prime}$. 
\label{NBexcess}}
\end{figure}

First, we describe the selection method of H$\alpha$ emitters at $z = 0.40$, 
which are basically similar to Fujita et al. (2003), Ly et al. (2007) and  Shioya et al. (2008). 
For emission-line galaxies at $z = 0.40$, redshifted H$\alpha$ emissions enter the NB921 band. 
We therefore select emission-line galaxies as NB921-excess objects. 
In order to select NB921-excess objects, we use $z^{\prime}$ band as off-band continuum. 
Taking account of the photometric error, we select NB921-excess objects using the following criteria: 
\begin{equation}
z^\prime - NB921 \ge \max[0.1, \; 3\sigma(z^\prime - NB921)], 
\end{equation}
where 
\begin{equation}
3\sigma(z^\prime - NB921) = -2.5 \log
(1-\sqrt{(f_{3\sigma_{NB921}})^2+(f_{3\sigma_{z^{\prime}}})^2}/f_{NB921}). 
\end{equation}
We show these criteria in Figure \ref{NBexcess}. 
In order to avoid the influence of saturation of brighter objects,
we adopt another criterion of $NB921 > 20$. 
For galaxies with $z^\prime > z^\prime_{\rm lim,3\sigma}$, we use the lower-limit 
value $(z^\prime-NB921)_{\rm low.limit} = z^\prime_{\rm lim,3\sigma} - NB921$ 
for our sample selection. 
We then select the NB921-excess objects to the $NB921_{\rm faintest}=25.63$ which 
is determined by $NB921_{\rm faintest} + 3\sigma(z^\prime-NB921)_{NB921 = NB921_{\rm faintest}} = z^\prime_{\rm lim,3\sigma}$. 
Although the median of the $z^\prime-NB921$ is slightly different from 0 
($\approx -0.05$, see Ly et al. 2007, for this offset), 
we do not apply any correction for the NB921 and $z^\prime$ magnitudes. 
We then find 2039 sources that satisfy the above criteria.
A narrowband survey of emission-line galaxies potentially detects galaxies 
with different emission lines at different redshifts.
The emission lines that can be detected in NB921 passband are 
H$\alpha$, H$\beta$, [O {\sc iii}] $\lambda \lambda$4959,5007, 
[O {\sc ii}] $\lambda$3727, Ly$\alpha$, and so on.
In order to distinguish H$\alpha$ emitters at $z = 0.40$ 
from such emission-line objects,
we investigate their broad-band colors comparing observed 
2039 emitters with model spectral energy distributions (Coleman, Wu, \& Weedman 1980).
In Figure \ref{Color}, we show the $B-R_{\rm c}$ vs. $R_{\rm c}-z^{\prime}$ color-color diagram 
of 2039 sources and the loci of model galaxies. 
We select H$\alpha$ emitters by using the following criteria, 
\begin{equation}
B-R_{\rm c} \ge 1.08 (R_{\rm c}-z^\prime) + 0.55 \; ,
\end{equation}
and
\begin{equation}
B-R_{\rm c} \ge 2.80 (R_{\rm c}-z^\prime) - 0.84 \; .
\end{equation}
To investigate how our selection criteria suffer from 
contamination of galaxies at different redshifts, we plot 
the colors of galaxies with a spectroscopic redshift 
[specifically, 28 H$\alpha$ emitters, 60 [O{\sc iii}] emitters, 24 H$\beta$ emitters
and 2 [O{\sc ii}] emitters presented in Cowie et al. (2004), and
2 H$\alpha$ emitter, 22 [O{\sc iii}] emitters and 4 [O{\sc ii}] emitters presented
in Ly et al. (2007)] in Figure 2. 
The contamination of [O{\sc iii}], H$\beta$ and [O{\sc ii}] emitters into H$\alpha$
is 1/112 ($< 1$\%). 
The contamination of H$\alpha$ emitters into [O{\sc iii}], H$\beta$ and [O{\sc ii}] 
emitters is 3/30 (10\%). 
These results are considered to justify our selection criteria. 
Finally, we select 356 H$\alpha$ emitter candidates.

We have already studied on H$\alpha$ emitters at $z=0.24$ in the SDF (Morioka et al. 2008). 
Using the SDF official photometric catalog, we have selected H$\alpha$ emitters at $z=0.24$. 
Our selection criteria are as follows: 
(1) $20 \le NB816 \le 26.1$, 
(2) $iz-NB816 > \max[0.1, 3\sigma(iz-NB816)]$, 
(3) $(B-V) > 1.6(V-R_c)-0.1$ \& $(B-V) > 3.1(V-R_c)-0.9$, and 
(4) $(B-V) > 0.8(R_c-z^\prime)+0.2$ \& $(B-V) > 2.5(R_c-z^\prime)-1.2$, 
where $iz$ continuum is defined as $iz = 0.57 f_{i^\prime} + 0.43 f_{z^\prime}$. 
In total, 258 H$\alpha$ emitters are selected. 

\subsection{H$\alpha$ Luminosity of H$\alpha$ emitters}

Since the clustering properties depend on the luminosity of galaxies, 
we have to know the luminosity range of our sample. 
In order to obtain the H$\alpha$ luminosity for each source, we correct for the presence of 
[N {\sc ii}] lines. Further, we also apply a mean internal extinction correction to each object.
For these two corrections, we adopt the flux ratio of 
$f(\textrm{H}\alpha) / f([\textrm{N {\sc ii}}] \lambda\lambda 6548,6584) = 2.3$ 
(obtained by Kennicutt 1992; Gallego et al. 1997; used by Tresse \& Maddox 1998;
Yan et al. 1999; Iwamuro et al. 2000) and $A_{\rm{H}\alpha}=1$ (Gallego et al. 1995). 
We also apply a statistical correction (28\% for NB921 and 21\% for NB816; 
the average value of flux decrease due to the filter transmission) 
to the measured flux of each objects because the filter transmission
function is not square in shape (Fujita et al. 2003).
The H$\alpha$ flux is given from the observed flux, 
$f_{\textrm{obs}}(\textrm{H}\alpha + [\textrm{N {\sc ii}}])$, by:
\begin{eqnarray}
f_{\textrm{cor}}(\textrm{H}\alpha) = f_{\textrm{obs}}(\textrm{H}\alpha + [\textrm{N {\sc ii}}]) \times
\frac{f(\textrm{H}\alpha)}{f(\textrm{H}\alpha + [\textrm{N {\sc ii}}])} \times 
10^{0.4A_{{\rm H} \alpha}} \times C \; ,
\end{eqnarray}
where $C=1.28$ for NB921 and 1.21 for NB816. 
Finally, the H$\alpha$ luminosity is given by $L(\textrm{H}\alpha) = 
4\pi d_{\rm{L}}^2 f_{\rm{cor}}(\rm{H}\alpha)$
where $d_{\rm L}$ is the luminosity distance at the redshift corresponding to 
the center of the filter passband: $d_{\rm L} = 2.17$ Gpc for $z=0.40$ and 1.22 Gpc for $z=0.24$. 

Figure \ref{HaLF} shows the H$\alpha$ luminosity functions of our samples. 
The H$\alpha$ luminosity function at $z=0.24$ is considered to be complete 
between $\log L({\rm H}\alpha) = 39.4$ and $\log L({\rm H}\alpha) = 40.8$. 
On the other hand, that at $z=0.40$ is incomplete for $\log L({\rm H\alpha}) < 39.8$. 
To compare the clustering properties of H$\alpha$ emitters without the luminosity effect, 
we made a subsample with the same luminosity range [$39.8 \le \log L({\rm H}\alpha) \le 40.8$]. 
The numbers of H$\alpha$ emitters whose luminosity range from $\log L({\rm H}\alpha)=39.8$ 
to $\log L({\rm H}\alpha) = 40.8$ are 139 at $z=0.24$ and 228 at $z=0.40$. 
We also made another subsample with $\log L({\rm H}\alpha) > 40.8$ for H$\alpha$ emitters at $z=0.40$ 
which contains 126 H$\alpha$ emitters. 

\subsection{Spatial Distribution and Angular Two-Point Correlation Function}

Figure \ref{XY} shows the spatial distribution of our H$\alpha$ emitters in the SDF. 
Left panel shows that at $z=0.24$ and right panel shows that at $z=0.40$. 
Objects in the range of $39.8 \le \log L({\rm H}\alpha) \le 40.8$ are shown 
with larger filled circles. 
To discuss the clustering properties quantitatively, 
we derive the angular two-point correlation function (ACF), $w(\theta)$,
using the estimator defined by Landy \& Szalay (1993),
\begin{equation}
w(\theta) = \frac{DD(\theta)-2DR(\theta)+RR(\theta)}{RR(\theta)},
\label{two-point}
\end{equation}
where $DD(\theta)$, $DR(\theta)$, and $RR(\theta)$ are normalized numbers of
galaxy-galaxy, galaxy-random, random-random pairs, respectively.
The random sample consists of 100,000 sources with the same geometrical
constraints as the galaxy sample. 
The formal error in $w(\theta)$ is described by 
\begin{equation}
\sigma_w=\sqrt{[1+w(\theta)]/DD}
\end{equation}
(Hewett 1982). 
Because of the finite size of the survey, this estimate will be negative offset 
from the true $w(\theta)$, which is called the integral constant. 
We calculate the integral constant, $C=0.02$ using the following definition
(Roche et al. 2002; see also Kova{\v c} et al. 2007): 
\begin{equation}
C = \frac{\sum RR A_w \theta^{-\beta}}{\sum RR} \; .
\end{equation}
This value is small and we neglect it in further calculations. 
Figure \ref{ACFall} shows the ACF of whole sample of 356 H$\alpha$ emitters in the SDF.
The ACF is fit well by power law, $w(\theta) = 0.0047^{+0.0010}_{-0.0009} \theta^{-0.94 \pm 0.04}$. 
In this section, we use data points between $0.001^\circ$ and $0.1^\circ$ for the power-law fit. 

It is useful to evaluate the correlation length $r_0$ of the two-point 
correlation function $\xi(r) = (r/r_0)^{-\gamma}$. 
The correlation length is derived from the ACF through Limber's equation 
(e.g., Peebles 1980). 
Assuming that the redshift distribution of H$\alpha$ emitters is 
a top hat shape of $z = 0.401 \pm 0.010$, we obtain the correlation 
length of $r_0 = 1.51$ Mpc. 
The two-point correlation function for all H$\alpha$ emitters 
is written as $\xi(r) = (r/{\rm 1.51 Mpc})^{-1.94}$. 
We also evaluate the ACF and correlation function for both subsamples of H$\alpha$ emitters 
at $z = 0.40$. 
The ACF of luminous H$\alpha$ emitters [$\log L({\rm H}\alpha) > 40.8$] is fit by power law, 
$w(\theta) = 0.0080^{+0.0077}_{-0.0039}\theta^{-0.86 \pm 0.15}$ and 
the correlation function is $\xi(r) = (1/{\rm 1.86^{+1.42}_{-0.88}Mpc})^{-1.86\pm0.15}$. 
The ACF of low-luminosity H$\alpha$ emitters [$39.8 < \log L({\rm H}\alpha) < 40.8$] is fit by power law, 
$w(\theta) = 0.0078^{+0.0033}_{-0.0023}\theta^{-0.84\pm0.08}$ and 
correlation function is $\xi(r) = (1/{\rm 1.62^{+0.64}_{-0.50}Mpc})^{-1.84\pm0.08}$. 
For H$\alpha$ emitters at $z = 0.24$ ($39.8 < \log L[{\rm H}\alpha] < 40.8$), the ACF is 
$w(\theta) = 0.013^{+0.005}_{-0.004}\theta^{-0.89\pm0.07}$ and 
correlation function is $\xi(r) = (1/{\rm 1.88^{+0.60}_{-0.49}Mpc})^{-1.89\pm0.07}$. 
For H$\alpha$ emitters at $z=0.24$, we assume that the redshift distribution of them 
is a top hat shape of $z=0.242 \pm 0.009$. 
We show these ACFs in Figure \ref{ACFcompare}.

\section{DISCUSSION}

Figure \ref{r0z} shows the correlation length $r_0$ as a function of redshift. 
For comparison, we also show those at $z < 0.15$ in the 2dFGRS (Norberg et al. 2001) 
and at $0.2 < z < 1.5$ in the VIMOS-VLT Deep Survey (VVDS) (Meneux et al. 2006) 
together with our results of H$\alpha$ emitters at $z = 0.40$ and 0.24. 
We also show the correlation length of H$\alpha$ emitters at $z = 0.24$ in the COSMOS 
field (Shioya et al. 2008). 
The two-point correlation function of H$\alpha$ emitters at $z=0.24$ in the COSMOS field, 
$\xi = (r/1.87^{+0.21}_{-0.20}{\rm Mpc})^{-1.88\pm0.03}$, is very similar to that 
for our sample at $z=0.24$, although we note that their average $B$-band absolute magnitude, 
$\langle M_B \rangle = -18.5$, is 1.5 mag brighter than our sample. 

First, the correlation lengths of our H$\alpha$ emitters [$39.8 < \log L({\rm H}\alpha) < 40.8$] at $z = 0.24$ and 0.40 
are smaller than those of early- and late-type galaxies at $0.2 < z < 0.5$ derived by 
Meneux et al. (2006) from VVDS. 
Since the correlation length depends on the luminosity of galaxies, 
we need to make a fair comparison using absolute magnitudes in the same rest-frame wavelength. 
The mean $B$-band absolute magnitudes, $\langle M_B \rangle$, 
of our sample with $39.8 < \log L({\rm H}\alpha) < 40.8$ 
are $\langle M_B \rangle = -16.9$ and $\langle M_B \rangle = -17.0$ 
for $z = 0.40$ and $z = 0.24$, respectively. 
On the other hand, the mean $B$-band absolute magnitudes of Meneux's 
sample are $\langle M_B \rangle = -19.1$ and $-18.1$ for early-type 
and late-type galaxies, respectively. 
The smaller correlation length of our H$\alpha$ emitters is interpreted 
by their small luminosity. 

\begin{figure}
\begin{center}
\FigureFile(70mm,150mm,90){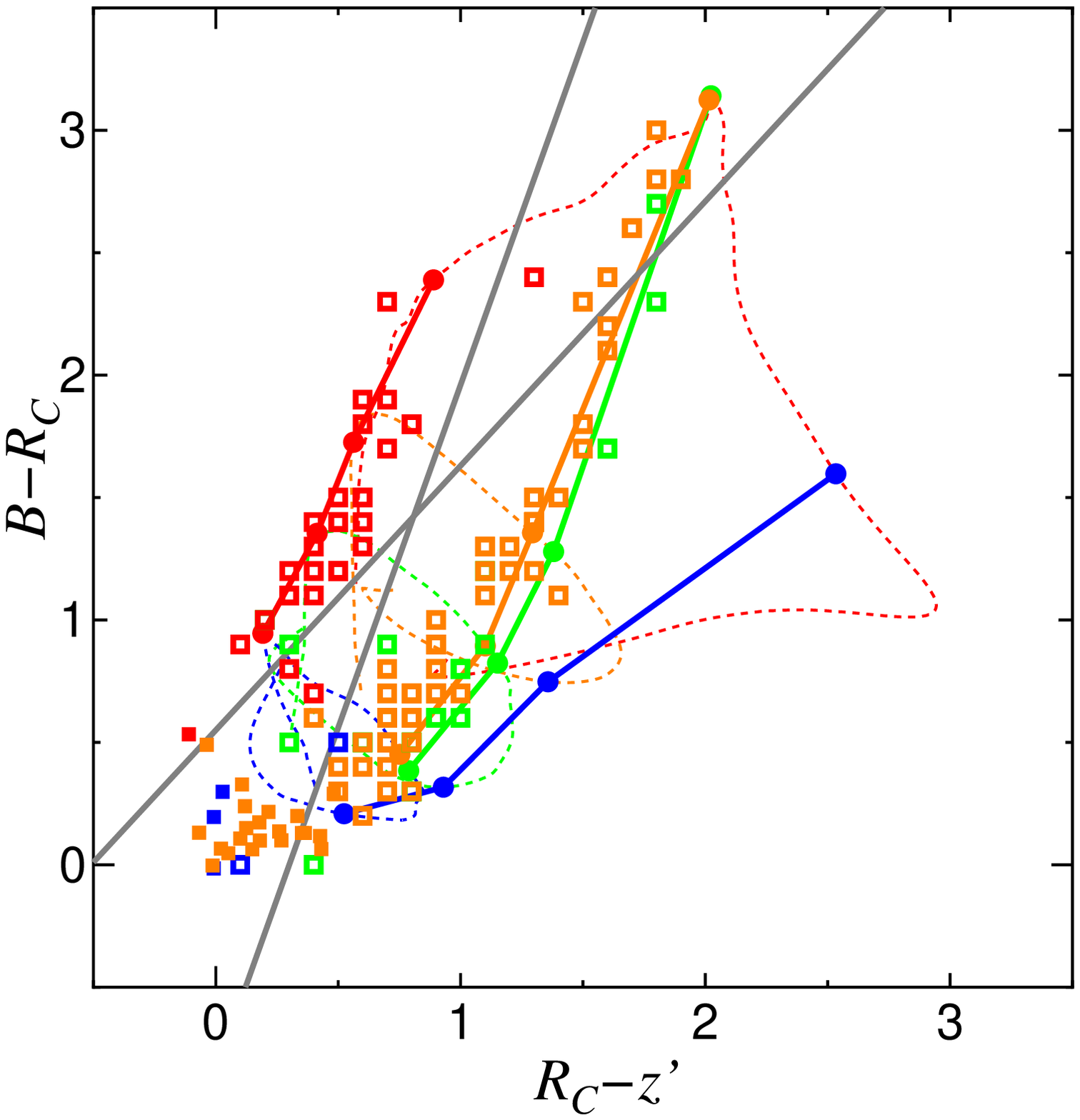}

\FigureFile(70mm,150mm){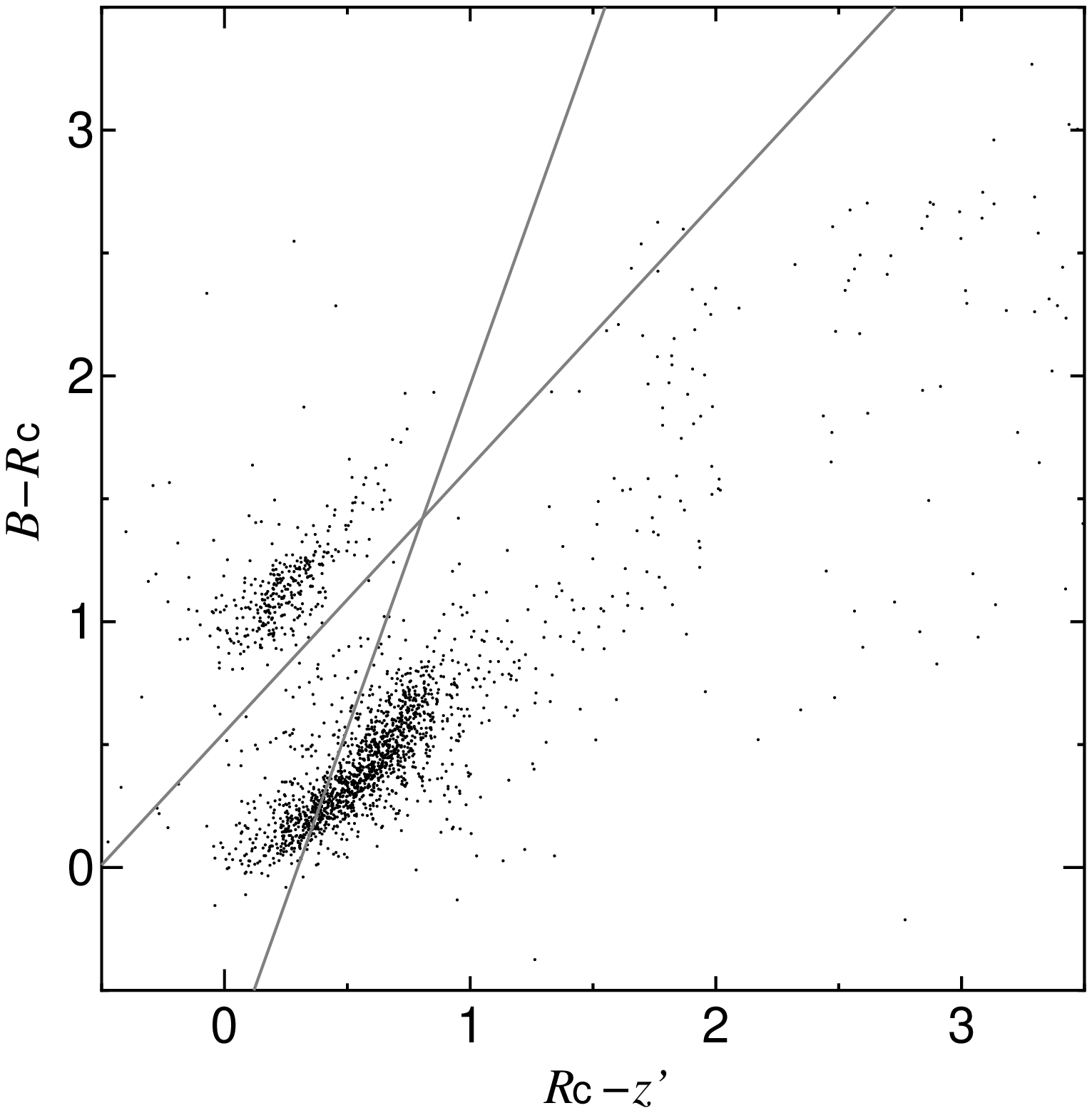}
\end{center}
\caption{
(Top) Color loci of model galaxies (CWW) from $z=0$ to $z=3$ are shown with 
dotted lines: red, orange, green, and blue lines show the loci of E, Sbc, Scd, Irr 
galaxies, respectively. 
Colors of $z=0.40$, $0.85$, $0.89$, and $1.47$
(for H$\alpha$, [O {\sc iii}], H$\beta$, and [O {\sc ii}] emitters, respectively) 
are shown with red, orange, green, and blue lines, respectively. 
Galaxies in the GOODS-N (Cowie et al. 2004) with redshifts corresponding to H$\alpha$ 
emitters, [O{\sc iii}] emitters, H$\beta$ emitters, and [O{\sc ii}] emitters are shown 
as red, orange, green, and blue open squares, respectively. 
Galaxies in the SDF (Ly et al. 2007) with redshifts corresponding to H$\alpha$ 
emitters, [O{\sc iii}] emitters, and [O{\sc ii}] emitters are shown 
as red, orange, and blue filled squares, respectively. 
(Bottom) Diagram between $B-R_{c}$ vs. $R_{c}-z^{\prime}$ for 
the 2039 emitter sources.
Our 356 H$\alpha$ emitters are isolated above the two gray solid lines.
\label{Color}}
\end{figure}

\begin{figure}
\begin{center}
\FigureFile(150mm,150mm){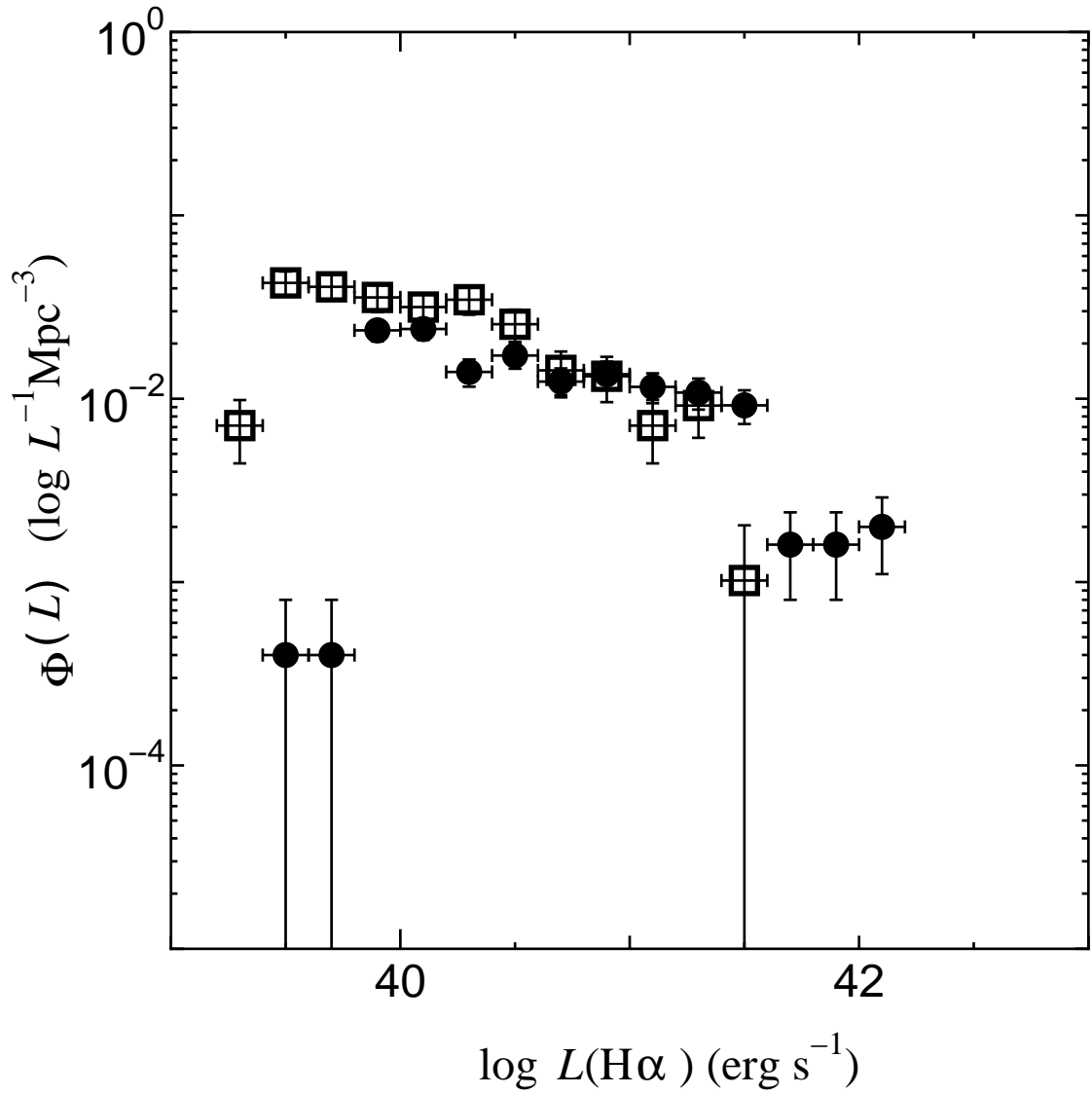}
\end{center}
\caption{
H$\alpha$ luminosity functions at $z = 0.24$ (open squares) and 
at $z = 0.40$ (filled circles). Since the H$\alpha$ luminosity function 
at $z = 0.24$ is considered to be complete at $39.8 < \log L({\rm H}\alpha) < 40.8$, 
we make subsamples of H$\alpha$ emitters both at $z = 0.24$ and $z = 0.40$ 
within the range of $39.8 < \log L({\rm H}\alpha) < 40.8$. 
\label{HaLF}}
\end{figure}

\begin{figure}
\begin{center}
\FigureFile(70mm,150mm){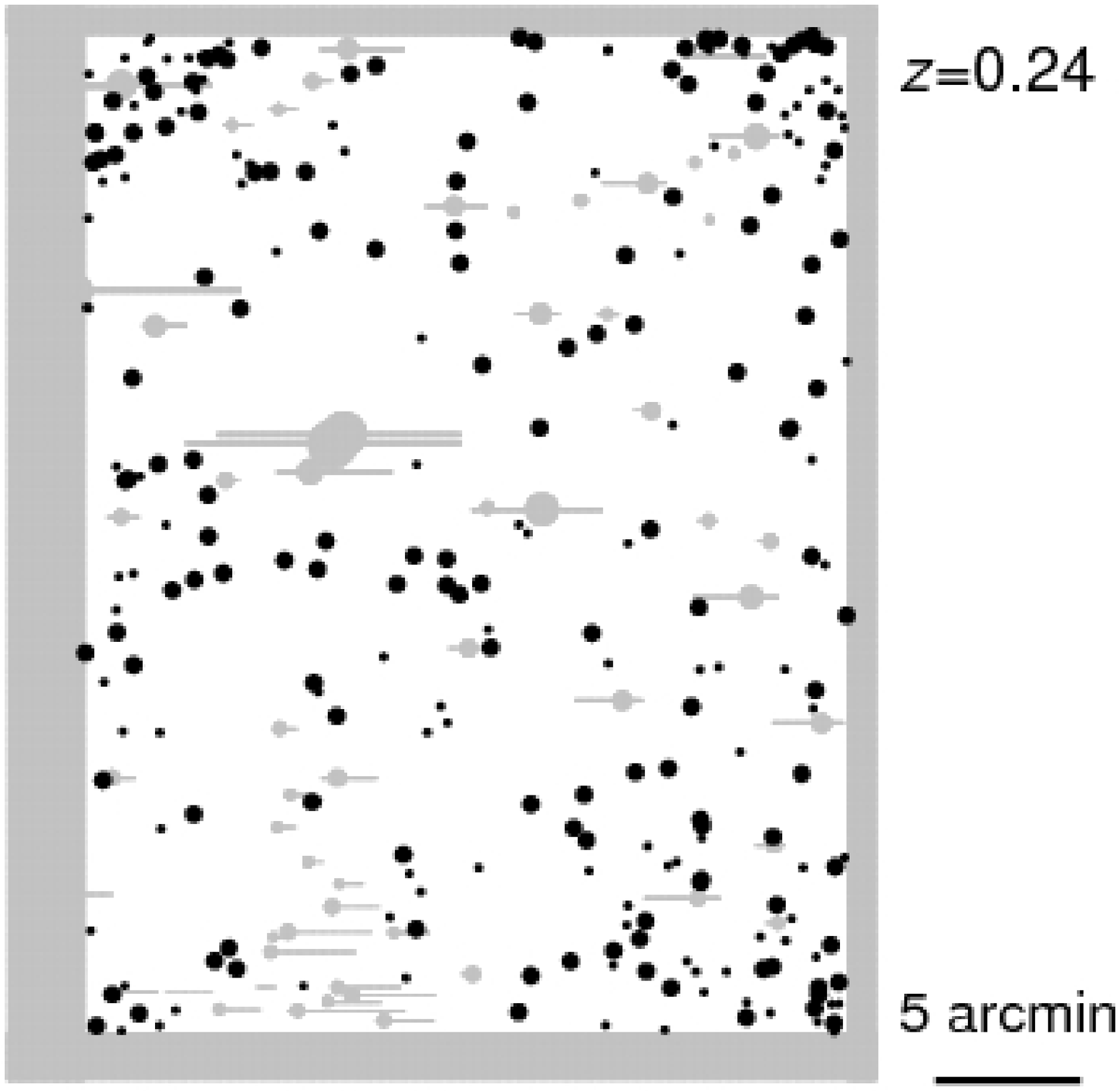}
\FigureFile(70mm,150mm){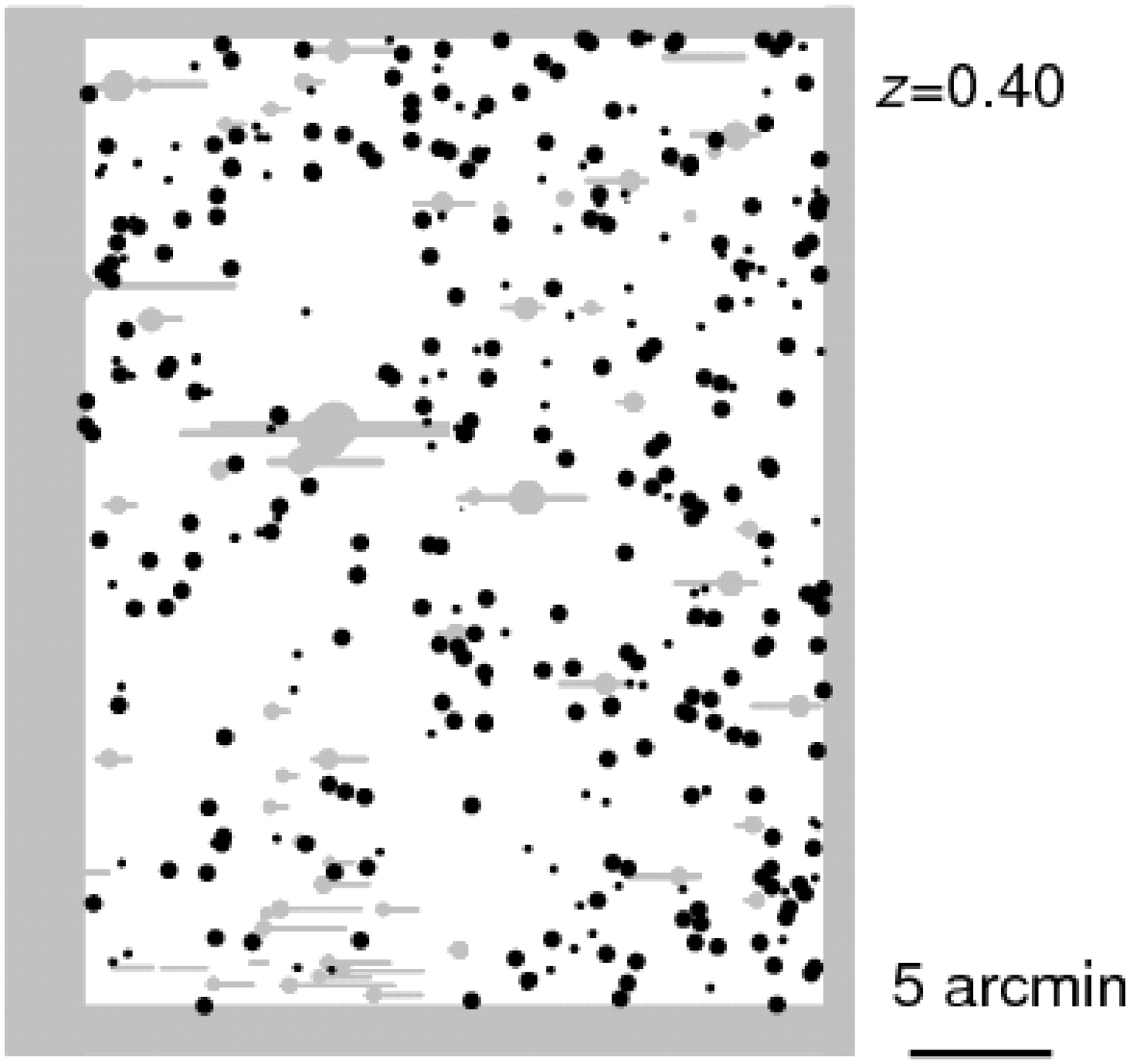}
\end{center}
\caption{
(Left) Spatial distribution of H$\alpha$ emitters at $z=0.24$. 
Large filled circles show objects in $39.8 \le \log ({\rm H\alpha}) < 40.8$. 
(Right) Spatial distribution of H$\alpha$ emitters at $z=0.40$. 
Large filled circles show objects in $39.8 \le \log ({\rm H\alpha}) < 40.8$ and 
small ones show objects in $\log ({\rm H\alpha}) > 40.8$. 
Shadowed regions show the areas masked out for the detection.
The field size of the SDF is $37^\prime \times 27^\prime$. 
\label{XY}}
\end{figure}

\begin{figure}
\begin{center}
\FigureFile(150mm,150mm){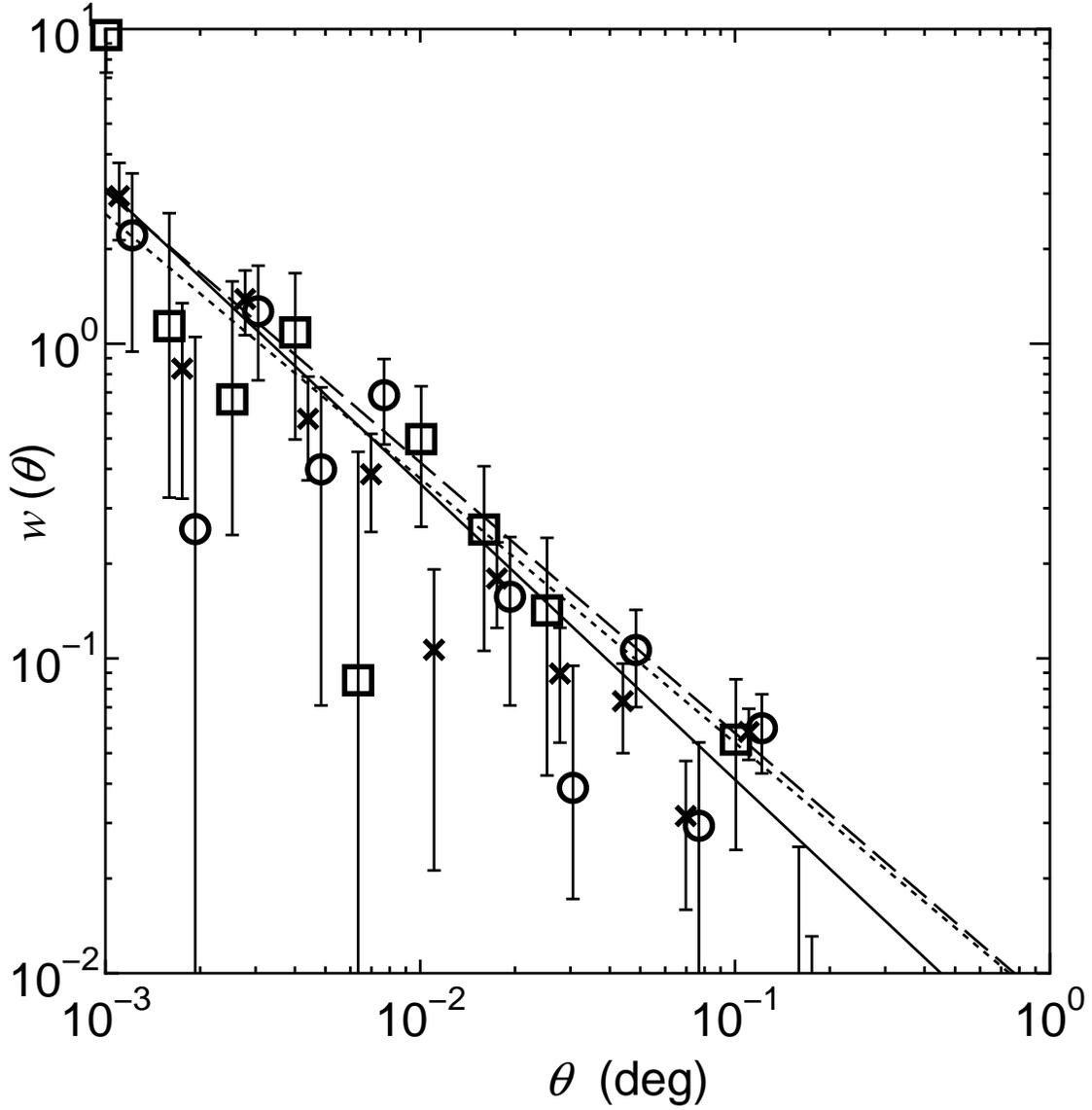}
\end{center}
\caption{
Angular two-point correlation function of all H$\alpha$ emitter candidates (crosses), 
the luminous subsample (open boxes), and the faint subsample (open circles) at $z = 0.40$. 
Solid line, dashed line, and dotted line show the best-fit single power law 
for all H$\alpha$ emitters, the luminous subsample, and the low-luminosity subsample, 
respectively. 
\label{ACFall}}
\end{figure}

\begin{figure}
\begin{center}
\FigureFile(150mm,150mm){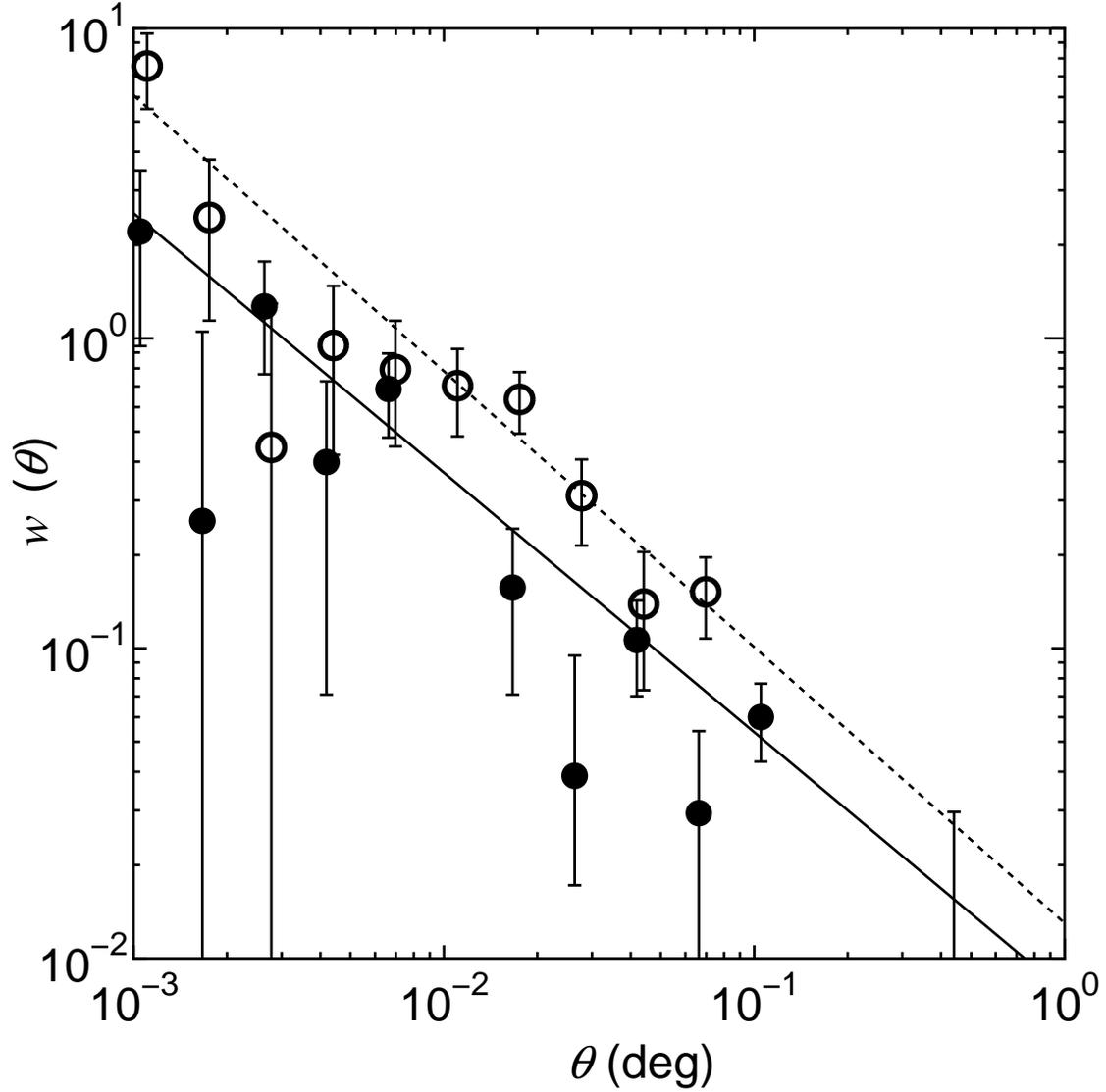}
\end{center}
\caption{
Angular two-point correlation function of H$\alpha$ emitter candidates of 
$39.8 < \log L({\rm H}\alpha) < 40.8$ (filled circles). 
Solid line shows the best-fit single power law: $w(\theta) = 0.0078 \theta^{-0.84}$. 
For comparison, we also plot the angular two-point correlation function of 
H$\alpha$ emitters at $z = 0.24$ of the same luminosity range (open circles). 
Dotted line shows the best-fit single power law: $w(\theta) = 0.013 \theta^{-0.89}$. 
\label{ACFcompare}}
\end{figure}

\begin{figure}
\begin{center}
\FigureFile(150mm,150mm){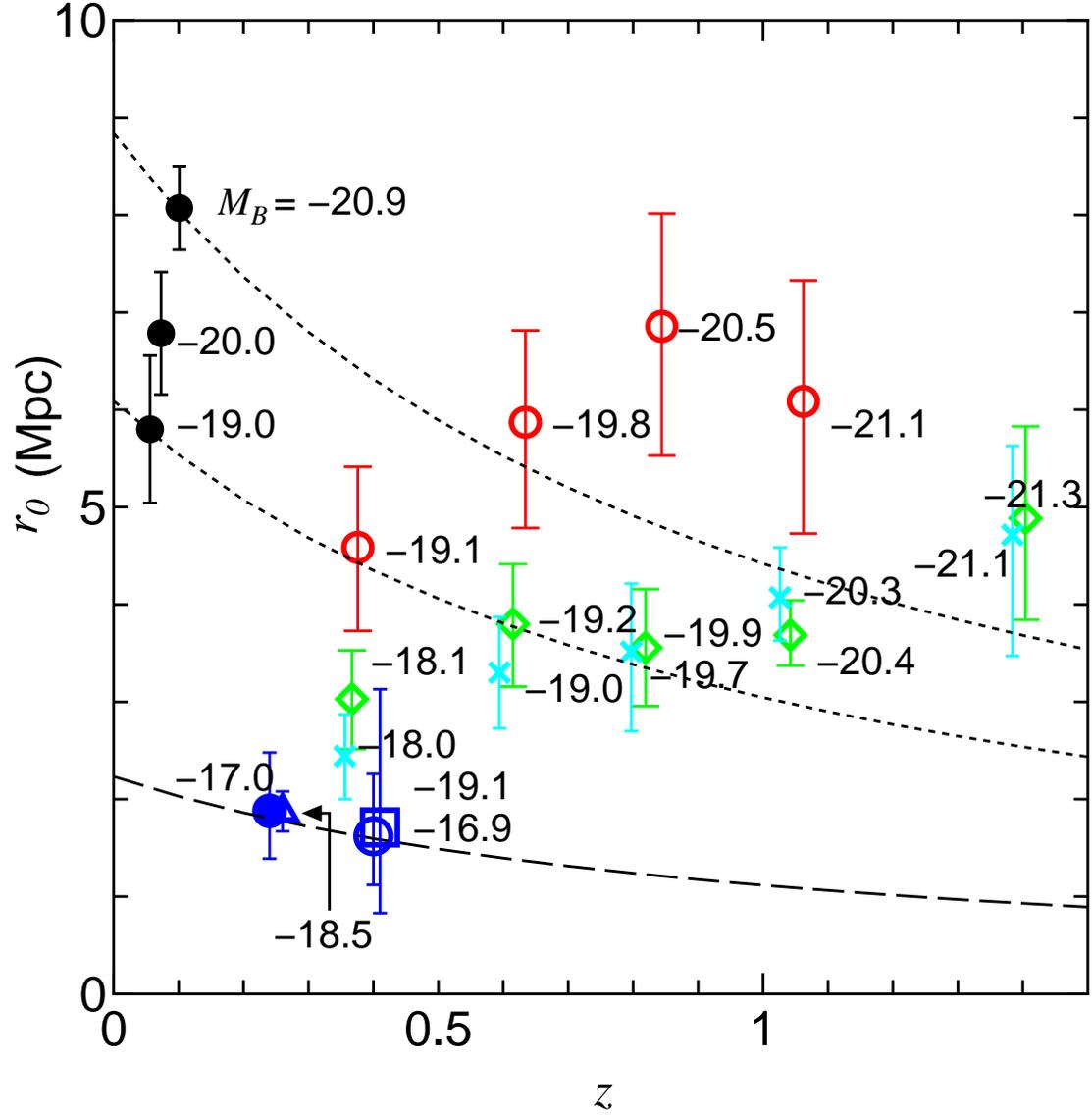}
\end{center}
\caption{
Correlation length $r_0$ as a function of redshift. 
Our results are shown as a blue open square (the luminous subsample at $z = 0.40$), 
a blue open circles (the low-luminosity subsample at $z = 0.40$), 
and a blue filled circle (H$\alpha$ emitters at $z = 0.24$). 
Blue triangle shows the correlation length for H$\alpha$ emitters at $z=0.24$ in 
the COSMOS field (Shioya et al. 2008). 
Red circles, green diamonds, and cyan crosses show 
early-type galaxies, late type galaxies, and starburst galaxies in Meneux et al. (2006).
Filled black circles show the $r_0$ at $z < 0.15$ for various $M_B$ (Norberg et al. 2002). 
The dotted and dashed curves are $r_0$ values as predicted by the ``$\epsilon$-model'' 
(Groth \& Peebles 1977) with $\epsilon = 0.8$ (corresponds to linear theory) and 
$\gamma=1.9$. 
The dotted curves show the predicted gravitational growth of the dark matter for 
galaxies with $M_B=-20.9$ and $M_B=-19.0$ of Norberg et al. (2001). 
The dashed curve show the predicted gravitational growth of the dark matter for 
our H$\alpha$ emitters at $z=0.24$. 
The number near each mark shows its absolute $B$ magnitude, $\langle M_B \rangle$. 
\label{r0z}
}
\end{figure}

We note that the clustering amplitude of our luminous H$\alpha$ emitter sample at 
$z = 0.40$ is weaker than those of the samples of Meneux et al. (2006). 
It may imply that the small clustering amplitude of our H$\alpha$ emitter sample 
depends not only on the low luminosity but also the spectral type, e.g., 
the clustering amplitude of red galaxies is larger than that of blue galaxies. 
As we shown in Figure \ref{Color}, the color of most of our sample is 
consistent with Scd - Irr type in Coleman et al. (1980), which is the same as 
type 3 \& 4 in Meneux et al. (2006). The clustering amplitude may depends on the 
equivalent widths of emission lines rather than the global SED. 
Our sample is biased toward galaxies with large emission-line equivalent width, 
which are considered to be in a very active phase of the star formation 
(e.g., Leitherer et al. 1999). 
We note that there is no significant difference in the correlation length between 
our two samples although it would be expected that the brighter sample could 
have a larger correlation length than that for the sample with 
$39.8 < \log L({\rm H}\alpha) < 40.8$. 
One possibility is that our brighter sample galaxies are due to brightening by 
the current star formation activity. 

Next, we compare the correlation length of the H$\alpha$ emitters at $z = 0.40$ 
with that of the H$\alpha$ emitters at $z = 0.24$. 
Although these are consistent within the error, the ratio of the $r_0$ at $z = 0.40$ 
to that at $z = 0.24$ (0.86) is also consistent with the expected value from the 
gravitational growth of dark matter halos. 
This fact implies that the H$\alpha$ emitters at $z = 0.40$ exist in the same environment 
where the H$\alpha$ emitters at $z = 0.24$ exist. 
If we assume the gravitational growth of dark matter halos to $z \sim 0$, 
the correlation length of this kind of galaxies evolve to $r_0 \sim 2.3$ Mpc. 
The ratio of the correlation length to that of the $L^*$ galaxies ($M^*=-20.5$) 
is 0.36. This ratio is smaller than the prediction of Norberg's law. 
It is suggested that the clustering amplitude of low-luminosity galaxies become smaller 
for lower luminosity galaxies. 

\vspace{1pc}
We would like to thank the Subaru Telescope staff for their invaluable help. 
We would like to thank Nobunari Kashikawa who is the chief of SDF,
and the other SDF members. 
This work has been financially supported in part by grants of JSPS (Nos. 15340059
and 17253001). 


\clearpage

\begin{table}
\caption{Best-fit parameters of Angular Correlation Functions.}\label{Ha:tab:parameters}
\begin{center}
\begin{tabular}{lccc}
\hline
\hline
{Sample} &
{$A_w$} &
{$\beta$} &
{$r_0$} \\
\hline
$z=0.40$ (all)                               & $0.0047^{+0.0010}_{-0.0009}$ & $0.94\pm0.04$ & $1.51^{+0.28}_{-0.28}$ \\
$z=0.40$ ($\log L({\rm H}\alpha)>40.8$)      & $0.0080^{+0.0077}_{-0.0039}$ & $0.86\pm0.15$ & $1.86^{+1.42}_{-0.88}$ \\
$z=0.40$ ($39.8<\log L({\rm H}\alpha)<40.8$) & $0.0078^{+0.0033}_{-0.0023}$ & $0.84\pm0.08$ & $1.62^{+0.64}_{-0.50}$ \\
$z=0.24$ ($39.8<\log L({\rm H}\alpha)<40.8$) & $0.013^{+0.005}_{-0.004}$    & $0.89\pm0.07$ & $1.88^{+0.60}_{-0.49}$ \\
\hline
\end{tabular}
\end{center}
\end{table}

\clearpage

\end{document}